\documentclass[utf8]{frontiersSCNS} 

\usepackage{url,hyperref,lineno,microtype,subcaption}
\usepackage[onehalfspacing]{setspace}
\usepackage{academicons}
\definecolor{orcidlogocol}{HTML}{A6CE39}

\def\keyFont{\fontsize{8}{11}\helveticabold }
\def\firstAuthorLast{Sun {et~al.}} 

\def\Authors{Tianrui Sun\,$^{1,2}$,
Lei Hu\,$^{1}$, Songbo Zhang\,$^{1}$,Xiaoyan Li\,$^{3}$, Kelai Meng\,$^{1}$, Xuefeng Wu\,$^{1,2,*}$, Lifan Wang\,$^{4}$, A. J. Castro-Tirado$^{5}$}
\def\Address{$^{1}$Purple Mountain Observatory, Chinese Academy of Sciences,Nanjing 210023, China\\
             
$^{2}$School of Astronomy and Space Sciences, University of Science and Technology of China, Hefei 230026, China \\

$^{3}$Nanjing Institute of Astronomical Optical Instruments, Chinese Academy of Sciences, Nanjing 210042, China \\   

$^{4}$ George P. and Cynthia Woods Mitchell Institute for Fundamental Physics \& Astronomy, Texas A. \& M. University, Department of Physics and Astronomy, 4242 TAMU, College Station, TX 77843, USA\\   

$^{5}$ Stellar Physics Department, Instituto de Astrofísica de Andalucía, Granada 18008, Spain
}
\def\corrAuthor{Corresponding Author}

\def\corrEmail{xfwu@pmo.ac.cn}

\begin{document}
\onecolumn
\firstpage{1}

\title[AST3-3 Pipeline]{Pipeline for Antarctic Survey Telescope 3-3 in Yaoan, Yunnan} 

\author[\firstAuthorLast ]{\Authors} 
\address{} 
\correspondance{} 
\extraAuth{}
\maketitle
\begin{abstract}
AST3-3 is the third robotic facility of the Antarctic Survey Telescopes (AST3) for transient surveys to be deployed at Dome A, Antarctica. 
Due to the current pandemic, the telescope has been currently deployed at the Yaoan Observation Station in China, starting the commissioning observation and a transient survey.
This paper presents a fully automatic data processing system for AST3-3 observations. The transient detection pipeline uses state-of-the-art image subtraction techniques optimised for GPU devices.
Image reduction and transient photometry are accelerated by concurrent task methods. Our Python-based system allows for transient detection from wide-field data in a real-time and accurate way. A ResNet-based rotational-invariant neural network was employed to classify the transient candidates. As a result, the system enables auto-generation of transients and their light curves.
\tiny
\keyFont{\section{Keywords:} data analysis, image processing, photometric
, transient detection, convolutional neural networks} 
\end{abstract}
\section{Introduction}
The Antarctic Survey Telescope (AST) 3-3 is the third telescope planned for time-domain surveys at Dome A, Antarctic. Before shipping to Dome A, it has been placed in Yaoan observation station of Purple Mountain Observatory for transient searching in the next several years. 
\cite{2015IAUGA..2256923Y} describes an overview schedule and designation for AST3 series telescopes. The AST3 series includes three large field-of-view (FoV) and high photometric precision 50/68 cm Schmidt telescopes \citep{2019RMxAC..51..135L}. 
The AST3-3 is designed for time-domain surveys in the $K$-band to search for transients in infrared at Dome A.
Due to the underdevelopment of infrared instruments of AST3-3, 
we temporarily use a CMOS  camera (QHY411 with Sony IMX411 sensor) with g-band filter for this commissioning survey instead.
This camera has an effective image area of 54 mm $\times$ 40 mm and a pixel array of 14304 $\times$ 10748 with exposure time ranges from 20 $\mu$s to one hour.
With the CMOS camera, the FoV is $1.65^{\circ}\times1.23^{\circ}$, the pixel scale is 0.41 arcsec, and the typical magnitude limit is $20\sim20.5$ in the $g$-band for 60s exposure images.

In the Yaoan observation station, we use the fully automatic AST3-3 telescope for a time-domain sky survey and follow-up observation.
We have constructed an observation scheme for the follow-up observation of transients according to the notices from  Gamma-ray Coordinates Network \citep{2008AN....329..340B}.
The summary of our observation system and hardware, the survey and target-of-opportunity strategy, and the early science results will be presented in a forthcoming publication (Sun et al. in prep.).

This paper gives a detailed overview of the AST3-3 data pipeline system. We have designed an automatic pipeline system containing data reductions, transient detection, and a convolutional neural networks (CNN) framework for transient classifications.
The data reduction pipeline includes instrumental correction, astrometry, photometry calibration, and image data qualification estimations. The transient detection pipeline consists of  the alignments of images, image subtractions, and source detection on subtracted images.

The image subtraction algorithm automatically matches the point spread function (PSF) and photometric scaling between the reference and science images.
In particular, one prevalent approach is the algorithm initially proposed by \cite{1998ApJ...503..325A}, and further developed by a series of works \citep{2000A&AS..144..363A,Bramich08,Becker12,Bramich13,2021arXiv210909334H}. This technique has been extensively used in the transient detection pipelines \citep[e.g.,][]{2015RAA....15..215Z, 2017PASA...34...37A, 2019PASP..131a8003M, 2020PASP..132l5001Z, 2022arXiv220102635B}.
In the last decade, it has played an important role in many successful time-domain survey programs, including intermediate Palomar Transient Factory \citep{iPTF_Cao16}, Dark Energy Survey \citep{DES18}, and  Panoramic Survey Telescope And Rapid Response System-1 \citep[PS1 hereafter,][]{PANSTARRS19}.

Time-domain surveys are demanded to find transients as fast as possible, but many bogus candidates are detected from the image subtraction results. The human workload can be greatly reduced by the classifier methods such as machine learning and deep learning \citep{2020MNRAS.499.3130G,2021Senso..21.1926Y}.
Random forest and some other machine learning methods previously attempted to solve the classification problem \citep{2015AJ....150...82G}.
We have applied a CNN framework to the estimation of the image qualification and candidate selections after transient extraction in this work. The CNN found optimal results with backpropagation \citep{1986Natur.323..533R}, and its accuracy has approached the human level in some classification and identifying tasks \citep{2015Natur.521..436L}.
Similar to some sophisticated neural networks, these CNN models can naturally integrate features from different levels and classify them in an end-to-end multilayer neural networks.
\cite{2015MNRAS.450.1441D} introduced the first rotation invariant CNN to classify galaxies by considering the inclination of the object in the classification.
The approach using rotation invariant CNN soon makes its way into the transient survey programs, e.g., The High cadence Transient Survey \citep[HiTS;][]{2016ApJ...832..155F} aimed at searching transients with short timescales.
In the transient detection procedure of the HiTS, \cite{2017ApJ...836...97C} used the rotation-invariant CNN to classify the real transient candidates and the fake candidates from the image subtractions. 
\cite{2019AJ....157..250J} modified the rotation-invariant CNN by adding the Long Short-Term Memory network \citep{HochSchm97} to enhance the performance in the satellite trail identifications. The alert classification system for the Zwicky Transient Facility survey also uses the updated rotation-invariant CNN \citep{2021AJ....162..231C}. 
The previous CNN structures for classification used superficial layers for feature extraction, and residual learning frameworks have been introduced in \cite{2015arXiv151203385H} to avoid the loss of too much information and the difficulty of deeper CNN.

In Section \ref{sect:level1}, we describe the data reduction pipeline and the qualification evaluation methods for image data. The transient detection pipeline for AST3-3 is presented in Section \ref{sec:transientpipe}. Section \ref{sec:CNN}  describes the CNN structure and training for classifying transient candidates and their performance. We show the conclusions in Section \ref{sec:conclusion}.

\section{Data Reduction Pipeline}
\label{sect:level1}
\begin{figure}[h!]
\begin{center}
\includegraphics[width=16cm]{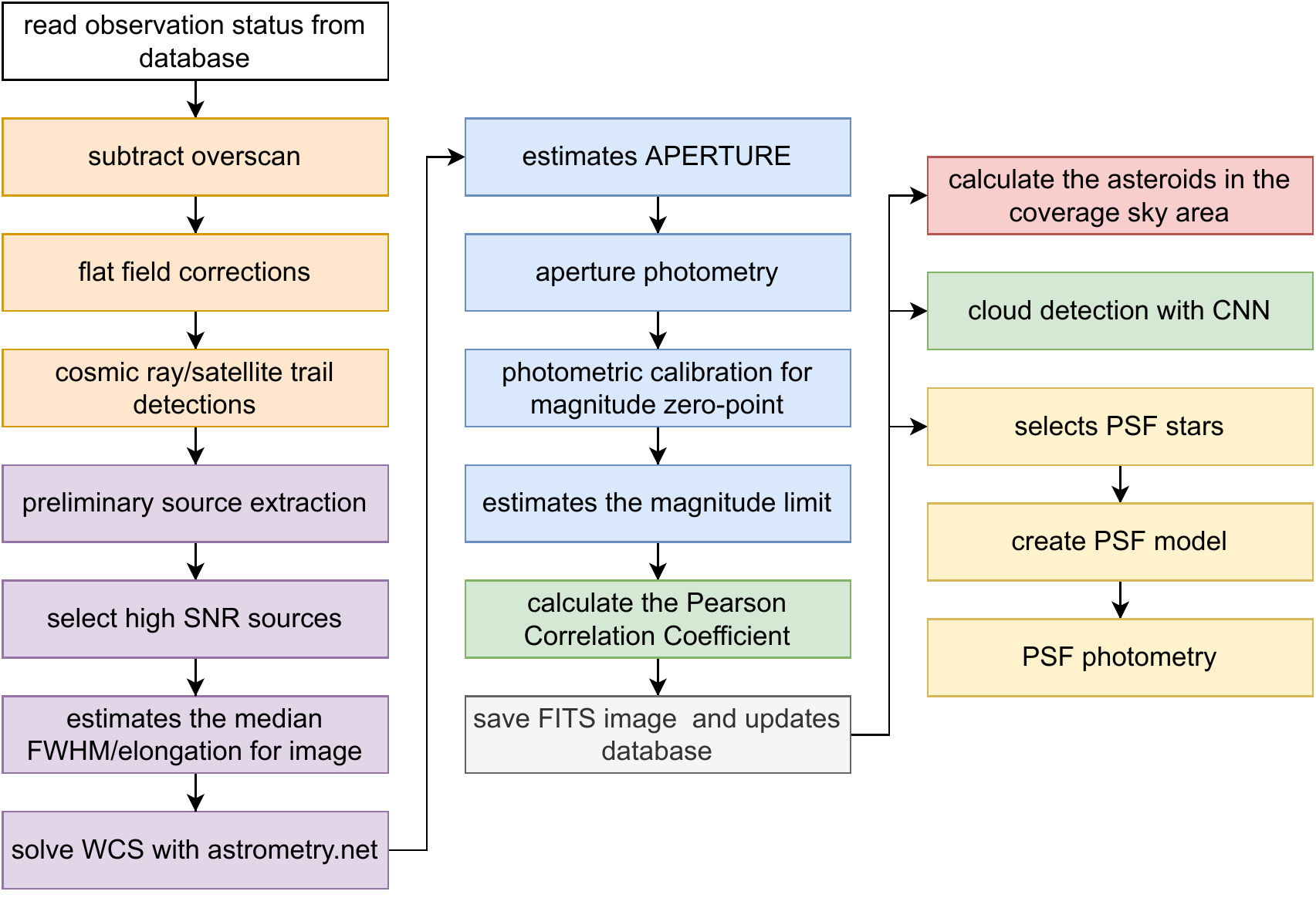}
\end{center}
\caption{The flow of the single-frame image processing for AST3-3. }
\label{fig:pipeline01}
\end{figure}

The data reduction pipeline aims at reducing the instrumental effects on the observational image to create the science image and apply the basic calibration information. This pipeline stage contains a group of subroutines for instrumental correction, astrometry, and photometry calibration. We also add a group of methods for evaluating the quality of images. The entire flow for the single-frame image processing shows in \autoref{fig:pipeline01}.

\subsection{Instrumental Correction}

The instrumental calibrations include overscan area removal and bias correction, flat-field correction, bad pixels, and cosmic-ray detection. The CMOS sensor of the camera on AST3-3 provides a 14304$\times$10748 pixel image array including a narrow overscan area of 50 lines. which contains the data for bias corrections. We use the median of the overscan area for each line as a bias field value similar to the CCDPROC method \citep{2015ascl.soft10007C} and produce an image array of 14206$\times$10654 pixels. The bias correction effectively removes the background boost from the offset value in the camera settings.

The AST3-3 telescope observes about 80 flat field images at a half-full maximum of the pixel capacity of near 30000 ADUs every observable twilight. The flat field image was constructed with the sigma-clipped median method like IRAF \citep{1999ascl.soft11002N} and applied to the observation images. The image also contains the cosmic ray defects on the detector, and we added the \textbf{L.A.Cosmic} package for the cosmic ray identification with the Laplacian edge detection method \citep{2001PASP..113.1420V}. A growing number of satellites plot trails on images that shall be added to the mask image. The pipeline uses the \textbf{MaxiMask and MaxiTrack} method for star trail detections \citep{2020A&A...634A..48P}.

\subsection{Astrometry Calibration}
\label{astrometry}
The pipeline for astrometry calibration solves the solution for the world coordinate system \cite[WCS,][]{2000ASPC..216..571C} and fits the WCS distortion parameters. 
The astrometry solution and the estimation of full-width half-maximum (FWHM) for images require source detection, and we optimised the sequence. 
The FWHM is a critical parameter in describing how the turbulence of the atmosphere and the properties of the optical system affect the observations of point sources.
The pipeline uses the \textbf{SExtractor} \citep{1996A&AS..117..393B} at first to detect sources on the image and measure basic parameters of stars with the automatic aperture photometry method.
The preliminary photometry catalogue results contain some bad sources, and we clean out the bad detection with the following restrictions:

\begin{itemize}
\label{createria1}
\item Exclude the neighbouring, blended, saturated or corrupted stars by removing the sources with FLAGS larger than zero.
\item Exclude the sources with the automatic aperture flux parameter FLUX\_AUTO that is not zero, and the ratio of FLUX\_AUTO and FLUXERR\_AUTO larger than 20. 
\item Select the isophotal and automatic aperture result parameter MAG\_BEST lower than 99 to exclude the bad magnitude fitting.
\item Exclude the outlier of FWHM\_IMAGE and elongations of sources by sigma clip method with $3\sigma$ threshold.
\item Sort the catalogue by the star-galaxy classification CLASS\_STAR and remove the last 20\% of the catalogue.
\end{itemize}

The remaining catalogue contains the most of detected point sources. We use the median of the FWHM\_IMAGE for sources in the remaining catalogue as the FWHM of the image. 
The pipeline calls the solve-field program in Astrometry.net \citep{2010AJ....139.1782L} to fit the WCS for their flexible local index files built from the Gaia Data Release 2 \citep{2018A&A...616A...1G}. We provide the X, Y, MAG\_BEST list for point sources, the pointing RA and DEC, the search radius, and the estimated pixel scale for the solve-field program to speed up the searching and fitting procedure.

AST3-3 has a large FoV of $1.65^\circ\times1.23^\circ$ which requires distortion corrections in the WCS. The pipeline also calculates the WCS with a fourth-order simple imaging polynomial \cite[SIP,][]{2005ASPC..347..491S} distortion for the accuracy of the coordinates. The mean solving time for ordinary observation is approximately 1.7 to 2 seconds for a catalogue generated by the SExtractor. 

\subsection{Photometry}
The pipeline extracts sources and estimates their flux and magnitudes on the image. We have applied the aperture photometry and PSF photometry methods in pipeline.
The aperture for the aperture photometry is determined with the \autoref{eq:aper}. The $\text{A\_IMAGE}$ is the semimajor axis value in the catalogue what matches the selection criteria in \autoref{astrometry} and the default value $C=6.0$ as derived in \cite{2018A&C....22...28S}, 

\begin{equation}
\text{Aper}=C\times\text{median}(\text{A\_IMAGE})
\label{eq:aper}
\end{equation}

The aperture photometry is performed by the SExtractor with the DETECT\_THRESH of $2\sigma$ and the estimated aperture. 
The pipeline cleans the catalogue from aperture photometry with the same distilling criteria described in \autoref{astrometry}. 
The estimation of magnitude zero-point calculates the difference between the aperture photometry and a reference catalogue.
We have selected the PS1 \citep[]{2017AAS...22922303C} as the reference catalogue for magnitude calibration for its sky coverage and much better magnitude limits.
We use a $\chi^2$ minimisation method introduced in the PHOTOMETRYPIPELINE \citep{2017A&C....18...47M} to generate the magnitude zero-point as the \autoref{eq:zp},

\begin{equation}
\chi^2=\sum_i^N\frac{(m_{zp}-\chi_i)^2}{\sigma^2_{\chi,i}}\label{eq:zp}
\end{equation}

In \autoref{eq:zp}, the i-th parameter $\chi_i$ is the difference between the aperture magnitude and the g magnitude in PS1 for the i-th source in the catalogue for all N matched sources. The magnitude zero-point $m_{zp}$ is determined by minimising $\chi^2$ with an iterative process to reject outlier samples.

The magnitude zero-point calibration has some residual due to the spatial variation in the atmospheric extinction across the large FOV.
We use a quadratic form to fit the offsets between the reference catalogue and the zero-point calibrated aperture magnitudes. The 2D polynomial formula to fit the residuals is shown in \autoref{eq:deltaM}, introduced in \cite{2007MNRAS.375.1449I},

\begin{equation}
\Delta m(x,y)= c_0+c_1x+c_2y+c_3xy+c_4x^2+c_5y^2
\label{eq:deltaM}
\end{equation}

$\Delta m$ is the zero-point offset for a star, and $c_i$ are the polynomial coefficients for the equation. Given that the background noise dominates the flux uncertainty of the faintest sources, the pipeline estimates the limiting magnitude based on the sky background with the method of  \cite{2010ApJ...719..900K} with the root-mean-square of background by the Python library SEP \citep{2016JOSS....1...58B}. 

The PSF photometry requires a group of well-selected stars for the profile fitting.  We select the sources in the aperture photometry catalogue with the same criteria in \autoref{createria1} but limit the restriction of FWHM\_IMAGE and ELONGATION to $1\sigma$ to keep the best ones. 

The pipeline feeds the selected catalogue to PSFEx  \citep{2013ascl.soft01001B} to generate a position-dependent variations  PSF model. The SExtractor accomplishes the PSF photometry with the result of PSFEx. It takes a relatively long time for PSF fitting and the model fitting in PSF photometry. This part works in the background after the single-frame image process.

\subsection{Data Quality Inspection} 
This subsection introduces the image quality inspection method, which estimates how the cloud affects the observation image.
The magnitude limit is an excellent parameter for the qualification of an image. For a large FoV telescope, the image extinction and airmass may vary across the FoV of the instrument, especially at lower altitudes.
The magnitude limit may also vary across the FoV of the instrument, producing spatial variations of the limiting magnitudes.
In ideal observation conditions, we suppose that the extinction is consistent everywhere in the image, which causes all-stars to have the same magnitude difference between machine magnitude and reference catalogue. 
Small clouds also influence the photometry results due to their discontinuous extinction to the nearby image parts. In the affected parts of the image, the flux of stars and galaxies would be reduced by clouds more than in other regions. 
In addition to the effects of clouds and extinction, incorrect WCS fits resulting in false star catalogue matches can seriously affect the magnitude corrections.

Based on the corrected aperture magnitudes and reference magnitudes being equal within the margin of error, we calculate the correlation between them. 
The measurement uses the Pearson correlation coefficient \citep[also called Pearson's r, PCC hereafter,][]{doi:10.1098/rspl.1895.0041}. The pipeline also selects stars that match the criteria as described in \autoref{astrometry} and calculates the PCC as \autoref{eq:pearsonr},

\begin{equation}
r= \frac{\sum((m_a-\bar{m_a})(m_g-\bar{m_g}))}{\sqrt{\sum(m_a-\bar{m_a})^2\sum(m_g-\bar{m_g})^2}}
\label{eq:pearsonr}
\end{equation}

\begin{figure}[h!]
\begin{center}
\includegraphics[width=16cm]{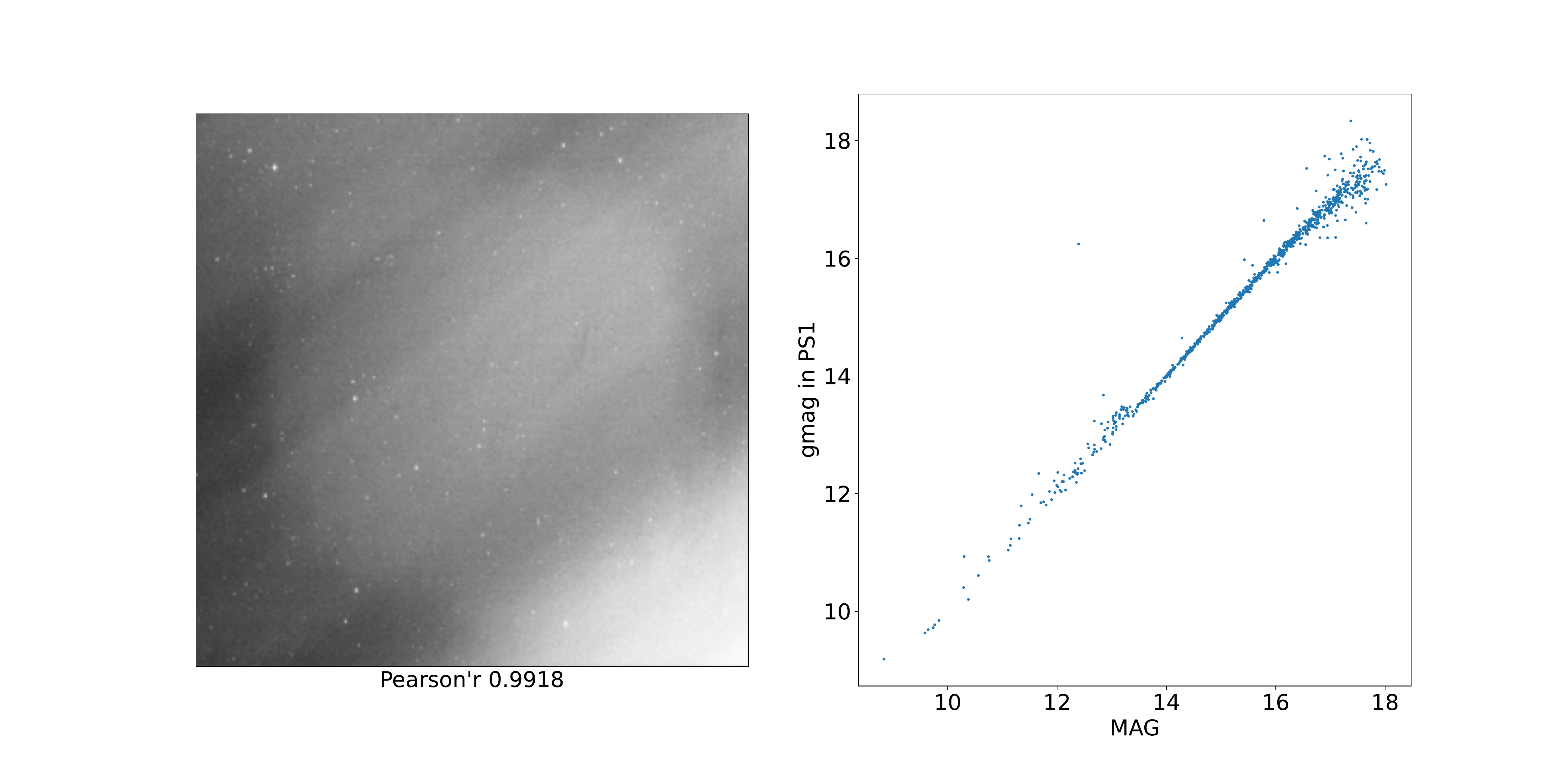}
\end{center}
\caption{The left panel shows the thumbnails of an AST3-3 image with obvious cloud effects. The right panel is a scatter plot of the calibrated aperture photometry magnitude and g-mag in PS1.}\label{fig:badpccexample}
\end{figure}

\autoref{eq:pearsonr} gives the expression of PCC where $m_a$ means the aperture photometry magnitude and $m_g$ is the $g$-band magnitude in PS1. For a completely ideal situation, the value of PCC becomes 1. For a typical AST3-3 image, the PCC value should be larger than 0.98 to avoid the influence of the clouds. 
When using PCC to estimate the image quality, we noticed that the uniform thin clouds do not influence the PCC in some images in a very significant way. \autoref{fig:badpccexample} shows a typical image of the good PCC with an obvious cloud on the image at an altitude of 53 degrees.

The discontinuous extinction would cause errors in flux calibrations in the image subtraction process for transient detection. As we can see the cloud from thumbnails in the left panel of \autoref{fig:badpccexample}, we could screen out the images with cloud effects manually. The pipeline creates thumbnails with the Zscale \citep{1999ascl.soft11002N} adjustment and normalisation method in Astropy \citep{2013ascl.soft04002G} to downscale it to an image size of 256 $\times$ 256 pixels.

Our manually checked results show that the cloud effects were still visible in the thumbnails. Thus, the cloud image classification for images is a simple classification problem that could be solved with the CNN method. 

\begin{table}[h!]
\begin{center}
\caption{Residual Neural Network Structure with the Residual Block}
\label{tab:cnnstructure}
\begin{tabular}{lccccc}
\hline
\textbf{Layer/Block} & \textbf{Input Channels} & \textbf{Output Channels}  & \textbf{Kernel Size}  & \textbf{Stride}   \\
\hline
Conv2d & 1 & 64 & $7\times7$ & 1 \\
BatchNormal2d & 64 & 64  & - & - \\
ResidualBlock & 64 & 64  & $3\times3$ & 1 \\
ResidualBlock & 64 & 64  & $3\times3$ & 1 \\
ResidualBlock & 64 & 128  & $1\times1$ & 1 \\
ResidualBlock & 128 & 128  & $3\times3$ & 1 \\
ResidualBlock & 128 & 256  & $1\times1$ & 2 \\
ResidualBlock & 256 & 256  & $3\times3$ & 1 \\
ResidualBlock & 256 & 512  & $1\times1$ & 2 \\
ResidualBlock & 512 & 512  & $3\times3$ & 1 \\
GlobalAvgPool2d & 512 & 512  & $4\times4$ & 3 \\
\hline
\end{tabular}
\end{center}
\end{table}

Consequently, we convert the cloud image classification into a simple image classification problem. 
We attempted to check for clouds in the images using the 18-layer residual neural nets \citep[ResNet-18, ][]{2015arXiv151203385H}, as shown in \autoref{tab:cnnstructure}.
The ResNet-18 is the simplest structure of residual neural networks, allowing a deeper network with faster convergence and easier optimization. 
We have chosen the original ResNet18 structure, as there are no vast data. The AST3-3 data are monochrome, which means we have only one channel of data to the input layer of the CNN. 
We have modified the input layer to input channel of 1, output channel of 64, kernel size of $7\times7$, and stride of 1 to match the thumbnail data and the second layer. 
Three fully-connected layers construct the classifier for the features extracted by ResNet-18. 
In the classifier part, we select the default rectified linear units \citep[ReLU, ][]{nair2010rectified} function as the formula of $\max(0,x)$ as the activation function.

The CNN training data set contains 1000 clear images and 1000 cloudy images as a balanced data set. 
To avoid overfitting in the CNN training, we use dropout to 0.5 in the classifier layers.
We have selected 20\% of the balanced data set as the test set for validation during the training of ResNet-18. The training uses the optimiser AdamW \citep{2017arXiv171105101L} and a scheduler for reducing the learning rate after each training epoch.
The CNN trained result shows accuracy on the test group of 98.35\% and a recall rate of 98.28\%. The CNN's classification of the thumbnail results is helpful as an indicator for significant cloud effects from our results. In practice, if CNN's classification of the thumbnail is cloud-affected, the image would be marked in the database and the website. If there is an obvious problem with the result of transient detection, like a massive number of detected candidates with a cloud-affected report, the image would be dropped automatically.

The data reduction pipeline schedules the single-image process independently to each image as a standalone thread concurrently.
The approximate average calculation time for the sparse starfield is 17 seconds per image after the observation. The pipeline also marks the necessary information into the FITS header and updates the database for the following step programs. 

\subsection{Data Reduction Pipeline Performance}
\begin{figure}[h!]
\begin{center}
\includegraphics[height=12cm]{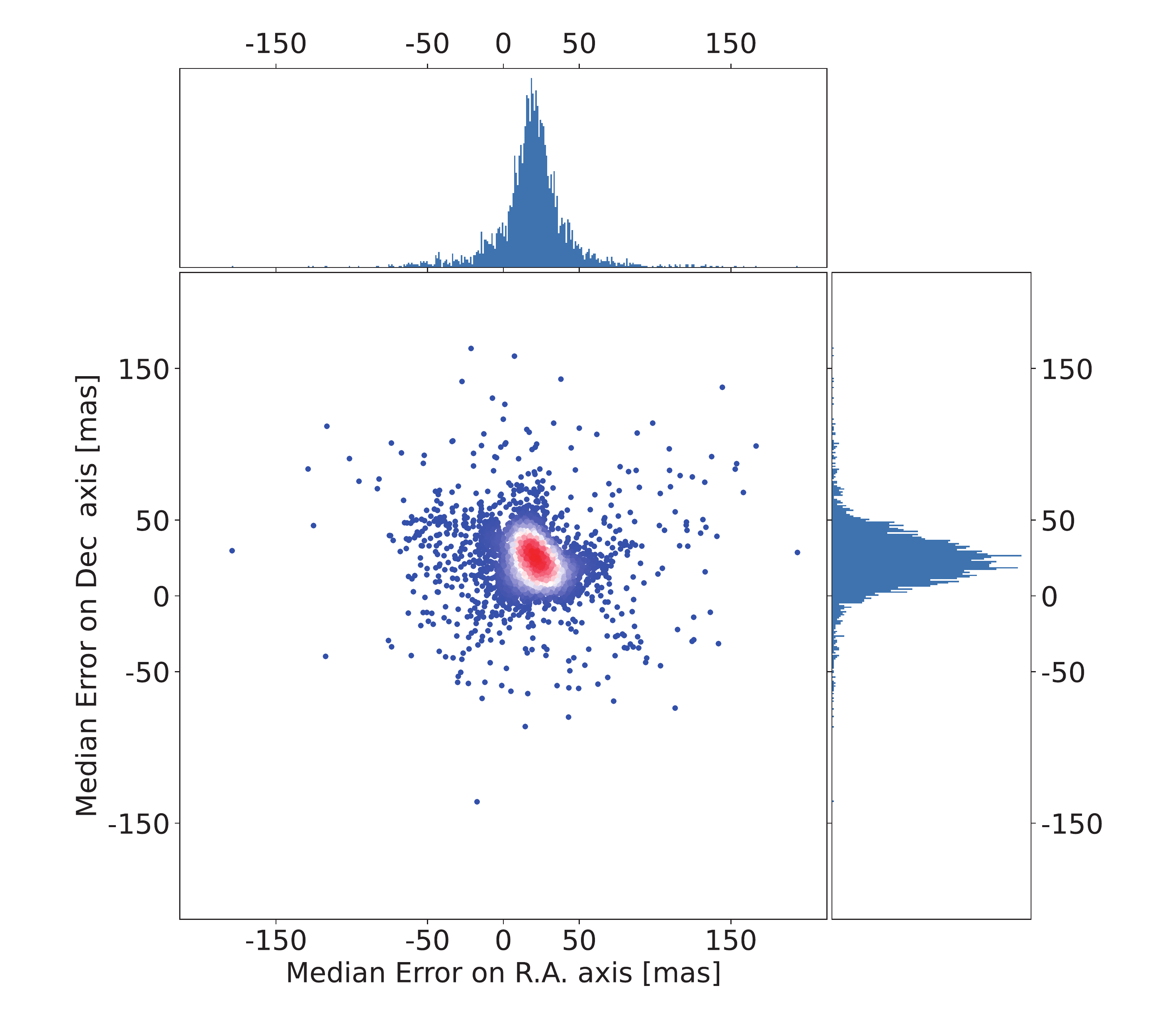}
\end{center}
\caption{
The main scatter-plot shows the median deviation distribution of the astrometric error along each axis with respect to the Gaia-DR2 catalogue for AST3-3 for stars in three thousand images. The upper and right panel shows the histograms of the distributions for both axes.
}\label{fig:wcsaccuracy}
\end{figure}

We have estimated the accuracy of the WCS for images by comparing the star position difference between the WCS  and the Gaia-DR2 catalogue. 
For each image, we match the catalogue between the aperture photometry result and the Gaia-DR2 catalogue and calculates the difference of right ascension and declination for the matched stars. The astrometry standard deviation in the data reduction pipeline is between 100 and 200 mas as shown in \autoref{fig:wcsaccuracy}.

\begin{figure}[h!]
\begin{center}
\includegraphics[width=15cm]{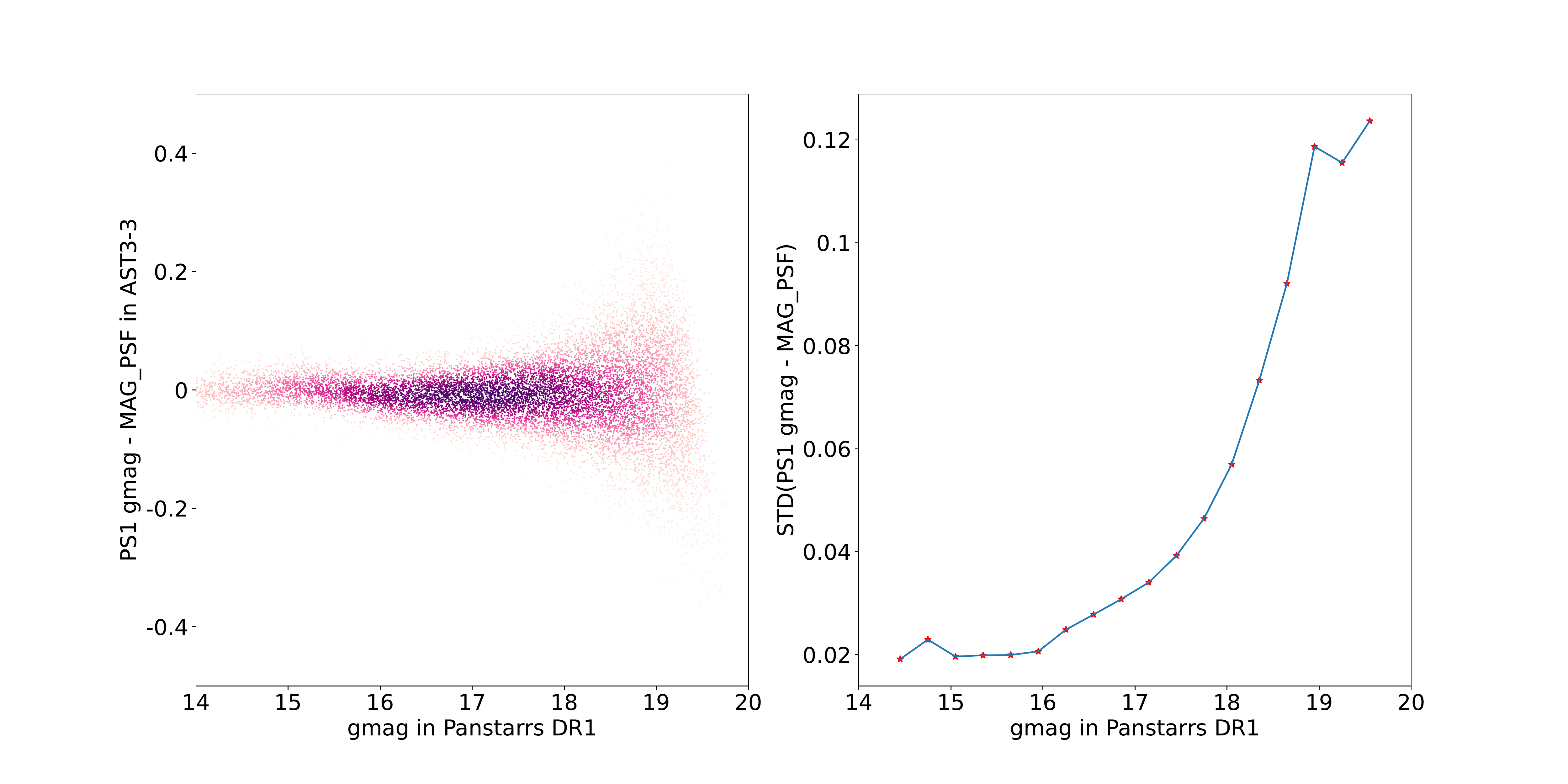}
\end{center}
\caption{
Left panel: The difference between the AST3-3 $g$-band photometry and PS1 $g$-mag values.
Right panel: The standard deviation of the binned and sigma clipped difference between the AST3-3 mags and PS1 $g$-mags.}\label{fig:magprecisi}
\end{figure}

The local PSF photometry results have some differences from the Pan-Starss DR1 catalogue due to some noise that should be well-fitted. We select an example of sky area `0735+1300' at an altitude 75 degrees above the horizon to show the result of PSF photometry fitting and the magnitude zero-point calibration. \autoref{fig:magprecisi} shows the difference map with density colour and the standard deviation trends from the detected bright to dark sources.

\begin{figure}[h!]
\begin{center}
\includegraphics[width=15cm]{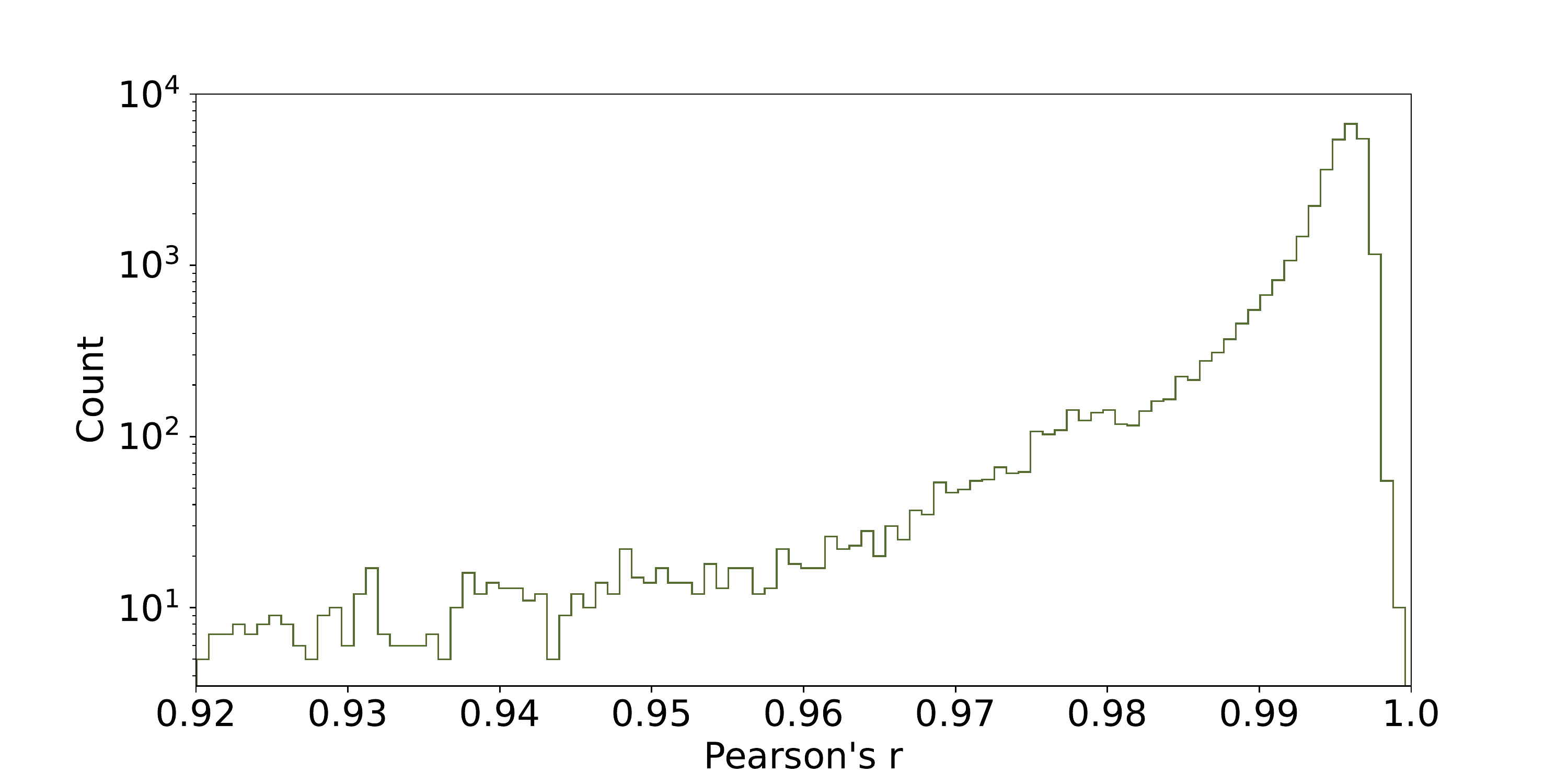}
\end{center}
\caption{The PCC value distribution for 33539 images since AST3-3 starts observation in 60 s exposure modes.}\label{fig:corr}
\end{figure}

\autoref{fig:corr} shows the PCC correlation distribution described in \autoref{sect:level1} since the first light in Yaoan observation station on 27 March 2021. In all the 60 s exposure images, 98\% of them had a PCC value greater than 0.95 and 84\% of them had a better PCC value of 0.99.

\section{Transient Detection Pipeline}
\label{sec:transientpipe}

The transient detection pipeline compares the newly observed science image with the previously observed template image data to find the transients appearing on images. 
This section builds a fully automatic pipeline for transient searching with the alignment method, image subtractions, and source detection on the difference images. 
We adopt the GPU version of the Saccadic Fast Fourier Transform algorithm \citep[SFFT hereafter,][]{2021arXiv210909334H} to perform the image subtraction.
The SFFT method is a novel method that presents the least-squares question of image subtraction in the Fourier domain instead of real space. 
SFFT uses a state-of-the-art $\delta$ function basis for kernel decomposition, which enables sheer kernel ﬂexibility and minimal user-adjustable parameters. 
Given that SFFT can solve the question of image subtraction with Fast Fourier Transforms, SFFT brings a remarkable computational speed-up of an order of magnitude by leveraging CUDA-enabled GPU acceleration. 
In real observational data, there are sources that can be hardly modelled by the image subtraction algorithm, and we should exclude them to avoid the solution of image subtraction being strongly misled. In our work, we use the built-in function in SFFT to pre-select an optimal set of sub-areas for a proper fitting.

\begin{figure}[h!]
\begin{center}
\includegraphics[width=15cm]{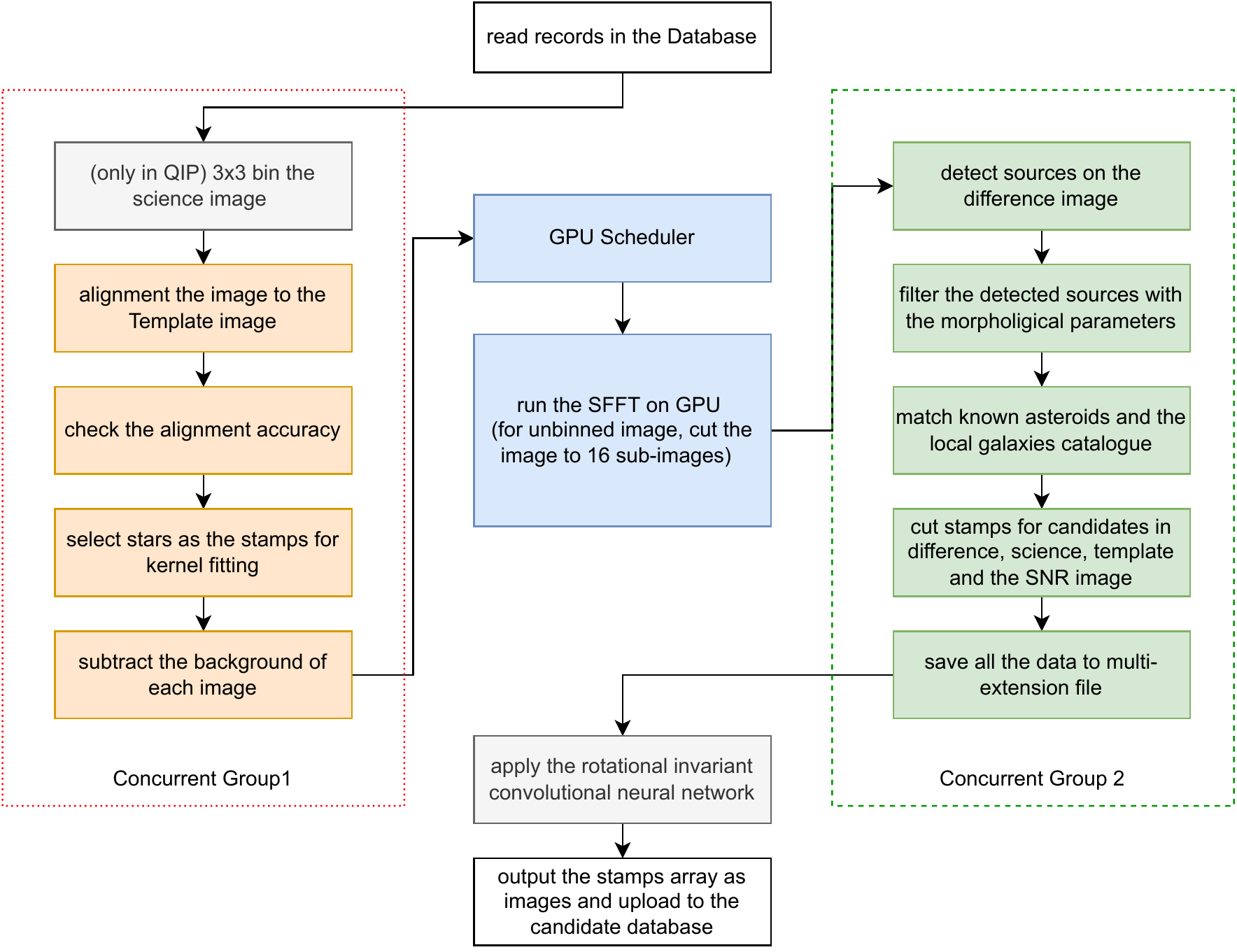}
\end{center}
\caption{This flowchart shows the transient detection pipeline for AST3-3. The pipeline is divided into four components. The left box shows the first concurrent group that contains the preparations for image subtractions. The middle part is the SFFT subtraction procedure, which calls the SFFT functions with a GPU resources scheduler. The right box shows the second concurrent group for candidate detection on the difference image. The bottom part shows the candidate classification and the data exchange of database. }\label{fig:pipelinediff}
\end{figure}

For the automatic follow-up observation of some transients, we need a rapid image subtraction process to search the possible transient candidates. AST3-3 science images are 14206$\times$10654 pixels in size, with a very sparse star field for high galactic latitude observations. The large data array takes a long time for kernel fitting and convolutions in image subtraction. 
We built two pipeline systems for transient detection. One focused on the quick analysis of the newly acquired image, especially for target-of-opportunity observations (quick image processing, QIP). And another one aimed at obtaining more reliable detections (deep image processing, DIP). The comparison of the two systems is shown in \autoref{tab:qipanddip}. 
QIP uses the 3 $\times$ 3 binned image of size 3552 $\times$ 4736 pixels to boost the image alignment and subtractions. The 3$\times$3 binned pixel scale increased to 1.23 from 0.41 arc sec per pixel, which increased the sky background noise. The flow chart of the pipeline system is shown in \autoref{fig:pipelinediff}

It is essential to prepare optimal template images to enable transient detections. 
We choose the earliest acceptable image taken for each sky area with restrictions on image qualities. The magnitude limit for templates should be better than 18.5 magnitudes. The PCC value is greater than 0.98, and there is no apparent cloud structure. The template image is copied directly after the data reduction pipeline result and tagged as the reference in the database. 
We create the template images for the QIP from the template image by the bin 3$\times$3 method. 
Since the template image for the QIP is a new image, we run the photometry and astrometry on the image with the same method in \autoref{sect:level1}.
The header of the WCS part of the template image is saved as a separate file to facilitate the SWarp \citep{2010ascl.soft10068B} for the alignment of the science image to the template. 

The strategies of each step for the QIP and DIP programs are the same: the resampling and interpolation of images for alignment, the kernel stamp selection, image subtractions and the source detections on the difference images. As a survey telescope, the AST3-3 has a fixed grid for observation. The shift and rotation between two images are tiny for each sky area but exist. Image alignment corrects the positional difference with a transformation matrix between the WCS of the template and science images. 
We use the SWarp for the resampling procedure with the LANCZOS3 algorithms. For the QIP part, the science image bin to 3552 $\times$ 4736 pixels by Scikit-image before resampling.

We perform image subtraction using SFFT for the binned images in QIP and the full-frame images in DIP branch, respectively. Note that the subtraction tasks are scheduled via database, and the two branches are triggered concurrently. The resulting difference images are used to detect transient candidates.

\begin{table}[h!]
\begin{center}
\caption{Processing Time for Different Methods}
\label{tab:qipanddip}
\begin{tabular}{lccccc}
\hline
\textbf{Pipeline} &\textbf{Method} & \textbf{Bin-Type}& \textbf{Image Size} & \textbf{Time} & \textbf{Stamp Size}\\
\hline
QIP& SFFT  & Bin3$\times$3 & 4736$\times$3552 & ~2s  & 31$\times$31\\
DIP& SFFT & Entire & 14206$\times$10654 & ~32s  & 91$\times$91\\
\hline
\end{tabular}
\end{center}
\end{table}

The calculations involved in SFFT are carried out using the multiple NVIDIA A100 GPUs equipped on our computing platform. For QIP case, we perform image subtraction straightforwardly for the 3$\times$3 binned images. 
For DIP case, we split the large full-frame image into a grid of sub-images with the size of 3072 $\times$ 4096 pixels to avoid memory overflow.

The pipeline uses the SExtractor to perform the target search on the difference images. We run the SExtractor with aperture photometry and threshold (DETECT\_THRESH) of $2\sigma$ for searching the star-like objects on the difference image. Some detected negative sources have obvious problems finding stars on the subtracted images. We cleared the bad sources with the following criteria:

\begin{itemize}
\label{transientdetectsx}
\item Exclude the sources with FWHM\_IMAGE lower than one pixel or more extensive than two times the image FWHM.
\item Exclude the sources with ELONGATION lower than 0.5 or larger than 6.
\item Exclude the sources with ISOAREA\_IMAGE less than 4.
\item Exclude the sources with FLAGS less than 4.
\item Exclude the sources near the image edge in 16 pixels.  
\end{itemize}

After the cleaning up, the catalogue of difference image becomes the candidate catalogue. Extragalactic transients have their host galaxies nearby, and most galaxies are already known. The pipeline cross-matches the candidate catalogue with the GLADE catalogue \citep{2021arXiv211006184D} with coordinate to obtain some near galaxy transients.

It is easy to detect many asteroids at Yaoan station due to its latitude. The pipeline calculates the location of asteroids by PyEphem \citep{2011ascl.soft12014R} with the Minor Planet Center Orbit (MPCORB) Database for each image. It also cross-matches all the candidate catalogues with the local asteroid catalogue to exclude the known asteroids.

The transient detection program pipeline ends when the provisional source detection is complete. 
As the transient detection program could re-produce all the data with the science images, there are retained for only 30-60 days to save storage space. 
We have built a multi-HDU (Header Data Unit) FITS format regulation to facilitate program error checking to store the data. The primary HDU storage only header information describing a summary of this image subtraction process results. The other HDUs store the subtracted image, the aligned image, the template image and the star list of the temporal source candidates through
compressed FITS image \citep{2009PASP..121..414P}.

\begin{figure}[h!]
\begin{center}
\includegraphics[width=14cm]{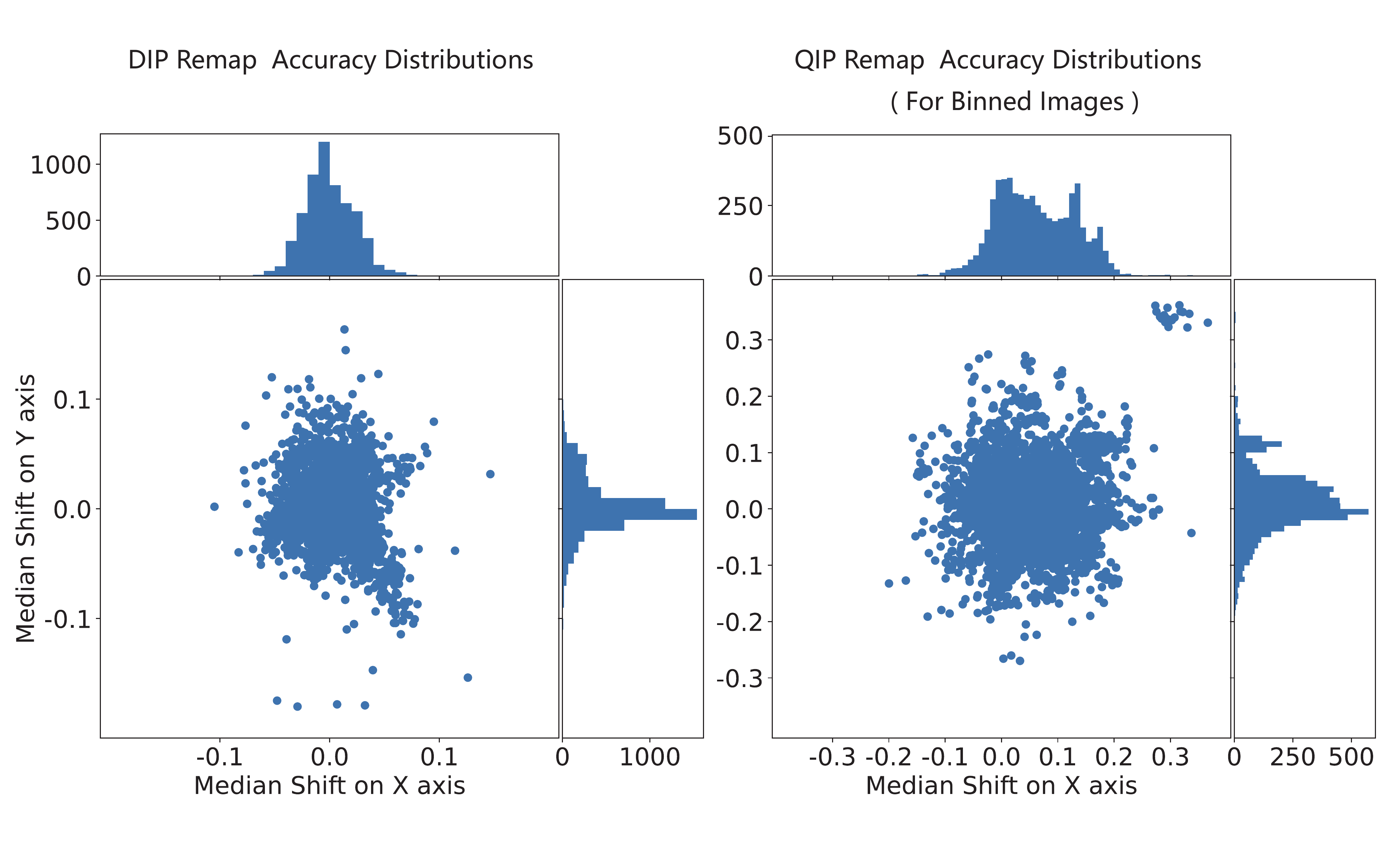} 
\end{center}
\caption{The two panels show the standard deviation of the x and y position differences in matched stars between the reference and remapped images. The right panel shows the unbinned images remap accuracy, and the left panel shows the remap accuracy for QIP.} 
\label{fig:remapstdFull}
\end{figure}

The image alignment is handled by SWarp based on the WCS information of the template and science images.
The position difference should be near zero for stars on both the template and alignment images to avoid the error occurrence during image subtraction. 
It's a fast method to check the accuracy of image alignment by examining the position difference between the catalogues from alignment image and the reference image.
Thence, we match the alignment catalogue by SExtractor and the reference catalogue with the grmatch program in FITSH  \citep{2012MNRAS.421.1825P}.
The position difference is calculated from the position of matched stars in the X, Y plane. \autoref{fig:remapstdFull} shows the median distributions of matched-star-position deviations of the DIP and QIP pipelines. The image alignment accuracy of our pipeline is typically less than 0.05 pixels, with a standard deviation of fewer than 0.2 pixels. The QIP have only slightly degraded accuracy due to the pixel scale binned to 1.23 arcsec.

\begin{figure}[h!]
\begin{center}
\includegraphics[width=15cm]{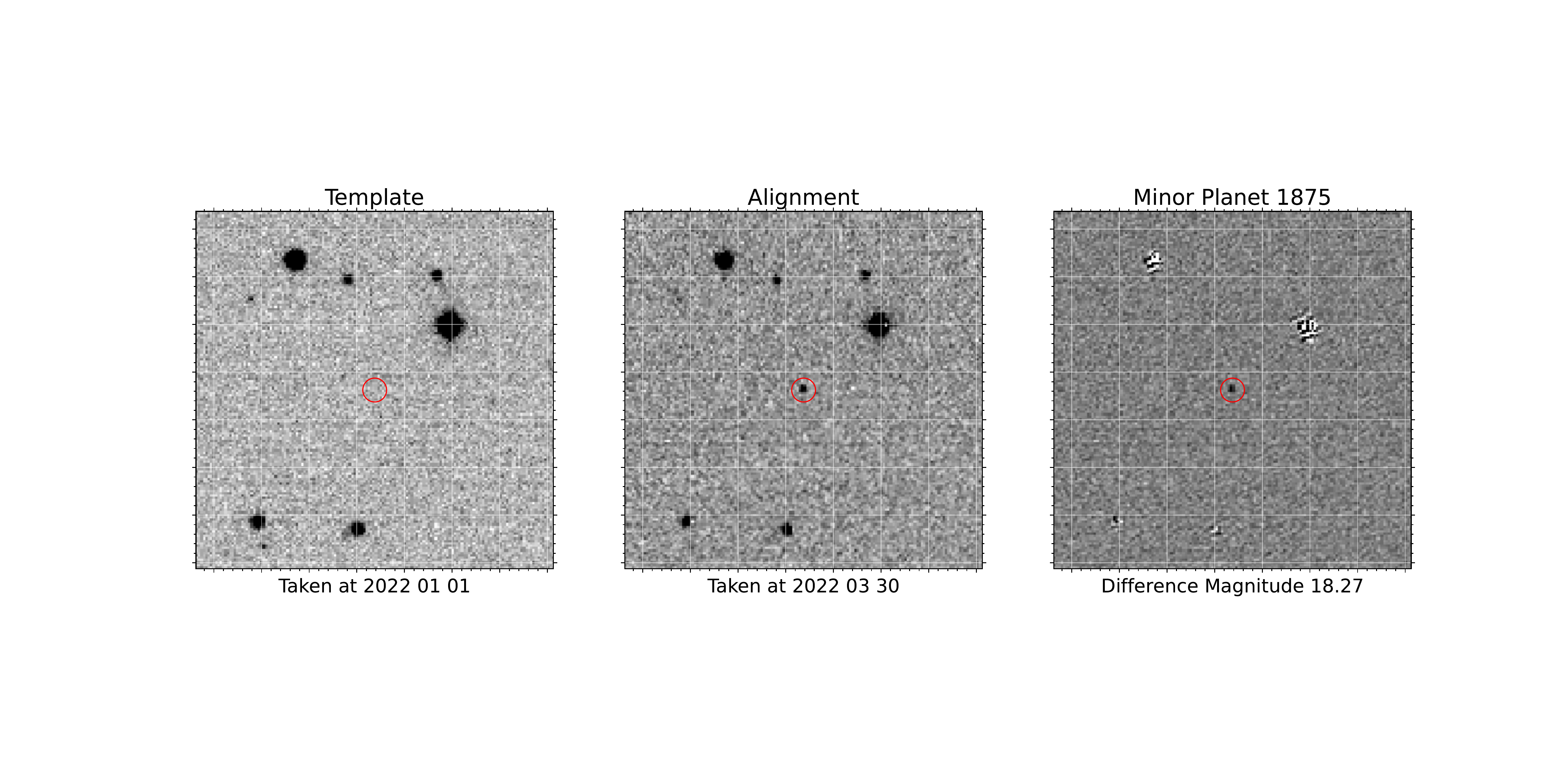} 
\end{center}
\caption{The example of the QIP: The detection of minor planet \# 1875 at magnitude 18.27. The left panel shows the template image taken at 1 Jan 2022 and the middle panel shows the remapped image taken at 30 Mar 2022. The right panel shows the difference image produced by the QIP pipeline.}\label{fig:MPC}
\end{figure}

The observation of asteroid 1875 is shown in \autoref{fig:MPC} by the time-domain survey and selected by the transient detection pipeline. It is a well-detected example of the pipeline described in this section. The target magnitude is 18.24 in an image with a magnitude limit of 19.9. The target can be seen clearly in the difference image and the pattern of nearby bright stars.

The performance of QIP and DIP procedures in images is only relevant to their pixel binning properties. As a result, the background and background's standard deviation increases, decreasing the limiting magnitude of the images in QIP. Since the QIP is only designed for the image with high priority observations, the resources used for QIP is restricted in both GPU time and the threshold of source extraction on the difference image. The QIP finished after the image was taken about 60 seconds, and the DIP finished after the image observation about at least 5-10 minutes due to the GPU time queuing.

\section{Candidates Classifications}
\label{sec:CNN}
The AST3-3 telescope is monochromatic in the $g$-band. It is difficult to distinguish among different types of transient sources with their morphological information in only several images. We divide the detected candidates into two categories: positive and negative candidates. The positive candidates are new point sources or variable sources on the science image. The positive candidates could be any astrophysical origin targets, while the negative candidates mainly originate from residuals and errors of the image subtraction pipelines. In this section, we choose to use the CNN-based approach to filter out the negative candidates from the image subtraction procedures.

\subsection{Rotation-Invariant Neural Network}
The original rotation-invariant CNN for classifying natural and artificial sources from transient detection pipelines is introduced in \cite{2017ApJ...836...97C} for HiTS as the CNN model named Deep-HITS. The Deep-HITS uses the data array by combining difference images, template images, science images and signal-to-noise ratio images as the stamps to classify the candidates generated from the detection after image subtraction. The original Deep-HITS network rotates the combination of images to 0, 90, 180, and 270 degrees to feed four independent CNN for feature extraction. The CNN parts of Deep-HITS only use a simple seven-layer structure that more complex neural networks can replace. 

It is expected that there should be nonlinear activation functions between the layers of CNN to enhance the performance of multilayer structures.
The rectified linear units \citep[ReLU, ][]{nair2010rectified} and its modifications are activation functions widely used in recent years. The ReLU function discards all negative values with $\max(0,x)$ for the sparsity of the network. The original Deep-HITS network uses the Leaky ReLU function, which improves the negative part by multiplication with 0.01 instead of zero as the formula $\max(0.01\times (x, x))$. In the development of Mobile-Net V3 \citep{2019arXiv190502244H}, the $\text{h-swish}$ function in \autoref{hswish} as the activation function to improve the accuracy of neural networks as a drop-in replacement for ReLU,

\begin{equation} \label{hswish}
\begin{split}
\text{ReLU6}(x) = \min(\max(0,x), 6) \\
\text{H-Swish}[x]=x\frac{\text{ReLU6}(x+3)}{6}
\end{split}
\end{equation}

\begin{figure}[h!]
\begin{center}
\includegraphics[width=5cm]{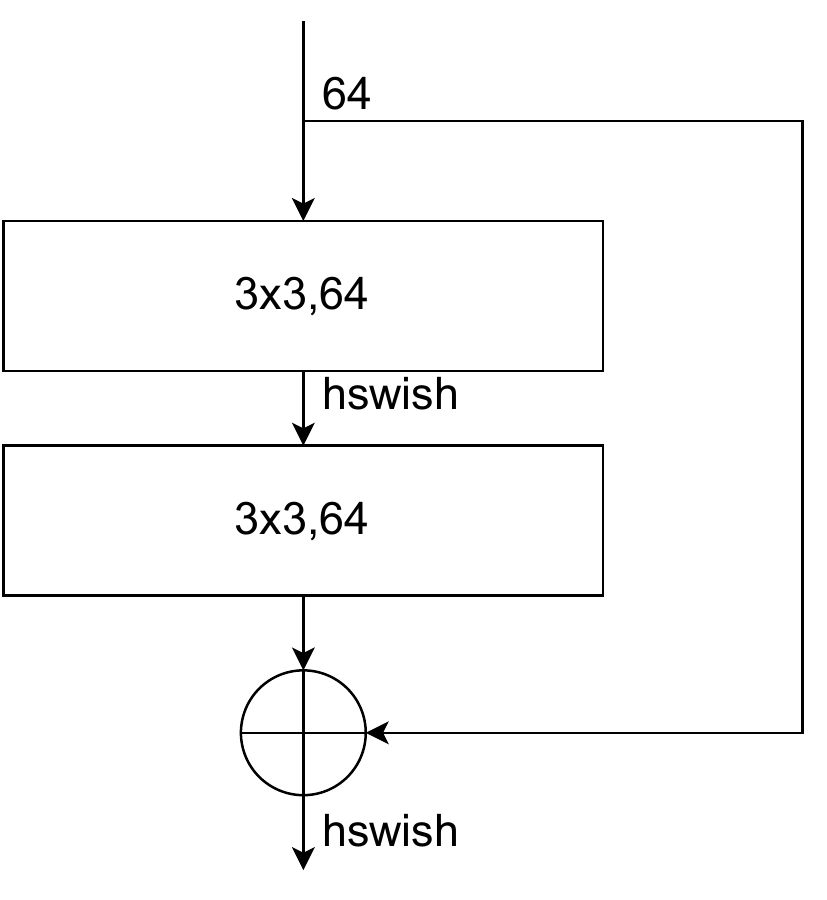}
\end{center}
\caption{
The Residual Block with the H-Swish Function.}\label{fig:residualbocks}
\end{figure}

\begin{figure}[h!]
\begin{center}
\includegraphics[width=10cm]{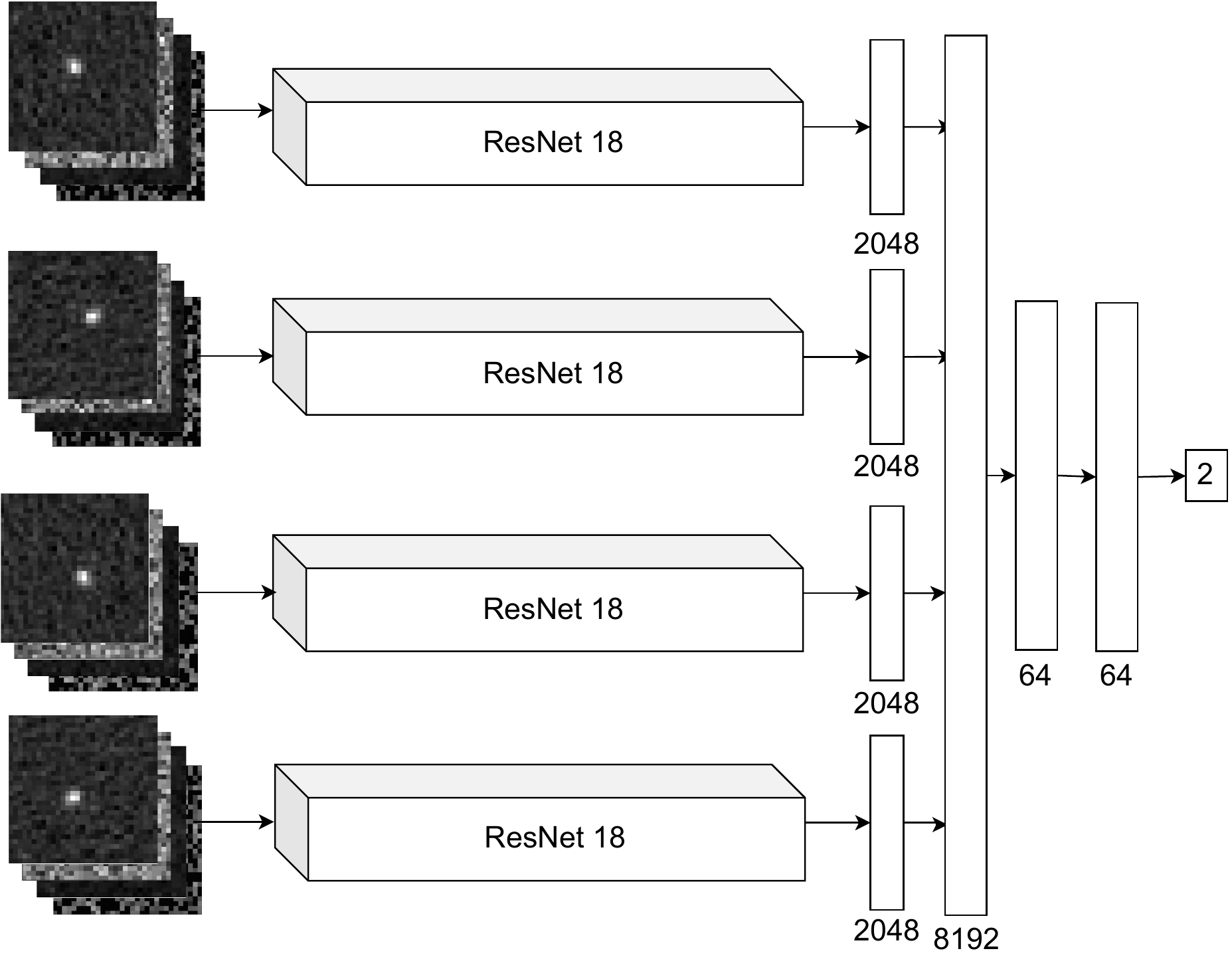}
\end{center}
\caption{
The Rotational Invariant Residual Neural Network}\label{fig:rires}
\end{figure}

We have modified the residual blocks in \autoref{fig:residualbocks} in the residual neural networks by changing the activation function to H-Swish. By organising the residual blocks to the structure of ResNet-18, we can replace the CNN parts of Deep-HITS with ResNet-18 as in \autoref{fig:rires}.

We build the three-dimensional array by combining each candidate's difference, template, science, and SNR stamps to feed the CNN models. For candidates from DIP, the array size is 91$\times$91$\times$4. For the candidates from QIP, the array size is 31$\times$31$\times$4 due to its 3$\times$3 binned image properties.
Before the CNN calculation, the pipeline rotates the stamp arrays 90, 180, and 270 degrees to feed the four branches of our modified Deep-HITS network. The main structure of our rotational invariant neural network is given in \autoref{fig:rires}. 

Our convolutional parts use the modified ResNet-18 for feature extraction as the structure \autoref{tab:cnnstructure}. The input channel is 4 to match the stamp arrays, and the output channel is 64 to feed the residual blocks. For the stamps with the 31$\times$31$\times$4 structure, the kernel size of the input layer is 3$\times$3, and the stride step is 1 in the QIP. For the stamps with 91$\times$91$\times$4 from the DIP, the first layer uses a 4$\times$4 kernel size and stride steps of 3 to match the residual block inputs. 

For each rotation of the stamp array, the modified ResNet-18 could create a vector of 2048 values as the feature extractions. We concatenate the output vectors of all rotations to a vector of $4\times2048$. The fully-connected layers are constructed by three linear layers and two H-Swish activation functions. These feature values were classified into two fully-connected layers. 

Due to the limitations of single-band observations, we only classify the candidates into positive and negative categories rather than performing a multicategory analysis. The classification of the results could be a binary problem that could use the simple cross-entropy loss function. The $p(x)$ in the function represents the true value $q(x)$  and represents the neural network classification value. The whole training problem of the neural network is thus to find the minimum cross-entropy under the training samples,
\begin{equation}
    H(p,q)= -\sum_x p(x)\log q(x)
\end{equation}

\subsection{Training and Accuracy}

The CNN training requires an extensive data set to avoid overfitting. The training data set should preferably be a balanced sample for CNN models. However, it is impossible to construct a balanced sample from the observational data. The negative candidates generated by transient detection are enormous, while the positive candidates are very rare in comparison. Thus, we use the simulated positive candidates to address this imbalance problem.

To create the simulated positive candidates, we select hundreds of science images with the best magnitude limits, lower airmass of observation and an excellent full-width half maximum of detected stars. For each image, the space-variation PSF is generated by PSFEx from the aperture catalogue with the same method described in the PSF photometry. 
The PSF model is constructed with stamp (VIGNET) size of 31$\times$31 and space-variation polynomial order of 3. We add the artificial stars to the selected images with magnitudes from 17.0 to 19.5 magnitudes at random positions. 
All positive candidates are selected with the human check and rejected for the wrong results.

\begin{figure}[h!]
\begin{center}
\includegraphics[width=12cm]{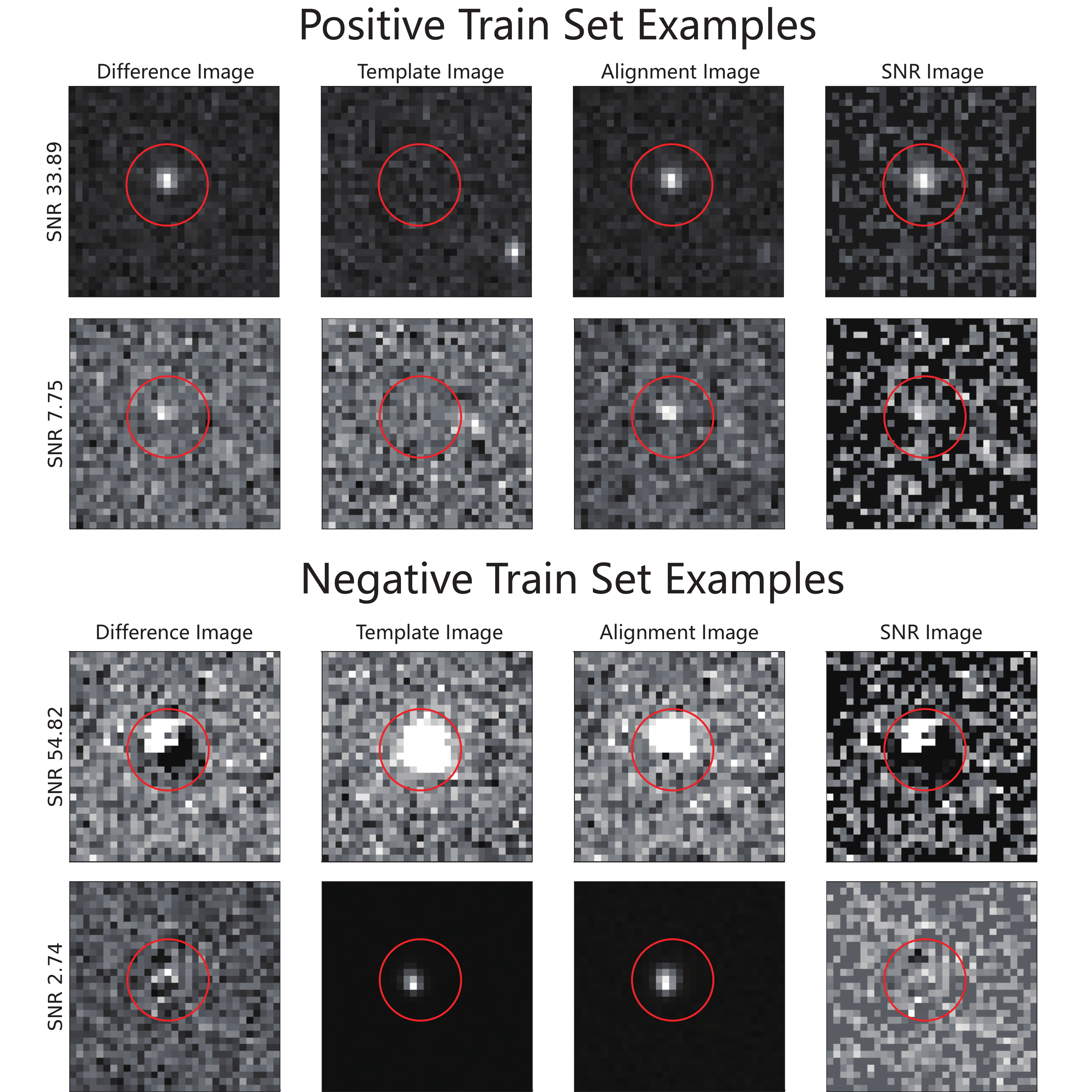}
\end{center}
\caption{
Examples of source samples for the positive and negative candidates used for training the neural networks. 
}\label{fig:trainsets}
\end{figure}

The negative candidates are bogus stamps caused by the residuals of alignments, isolated random hot pixels after resampling, cosmic rays, saturation stars, and some failed fitting sources. The negative candidates are produced from the image subtraction pipeline with real images and also selected by human. We cut the stamps for positive and negative candidates with the image subtraction pipeline. We constructed data sets with $10^4$ positive and $10^4$ negative candidates. Examples of the train set are shown in \autoref{fig:trainsets}.

Before the CNN training, we split 20\% candidates into the test group for training and validation groups. The rotational invariant residual neural network trains with the optimiser AdamW, batches of 256 stamp arrays and a dropout rate of 0.5 in the fully-connected layer. 
The learning rate is set to a relatively low value of 0.01 at the beginning, and it steps down by multiplying by 0.9 after each epoch. 
We build the neural network with the PyTorch \citep{NEURIPS2019_9015} and train it on the Nvidia A100 graphic processor unit. The training of our model requires a huge GPU memory, especially for the stamp size of 91$\times$91 for the stamps from the DIP pipeline. The accuracies and precisions of the CNN models for QIP and DIP programs are shown in \autoref{tab:trainresult}.

\begin{table}[h!]
\begin{center}
\caption{Precision of the CNN Model for stamps from QIP and DIP}
\label{tab:trainresult}
\begin{tabular}{lcccccc} 
\hline
\textbf{Method} & \textbf{Stamp Size} & \textbf{Accuracy}  & \textbf{Recall}  & \textbf{Precision}  & \textbf{F1Score}  \\
\hline
Bin3$\times$3 & 31$\times$31 & 99.88\% & 99.87\% & 99.87\% & 99.88\%\\
Unbinned & 91$\times$91 & 99.20\% & 99.20\% & 99.21\% & 99.20\%  \\
\hline
\end{tabular}
 \end{center}
\end{table}

\subsection{Rotational-Invariant Residual Neural Network Performance}

\begin{figure}[h!]
\begin{center}
\includegraphics[width=15cm]{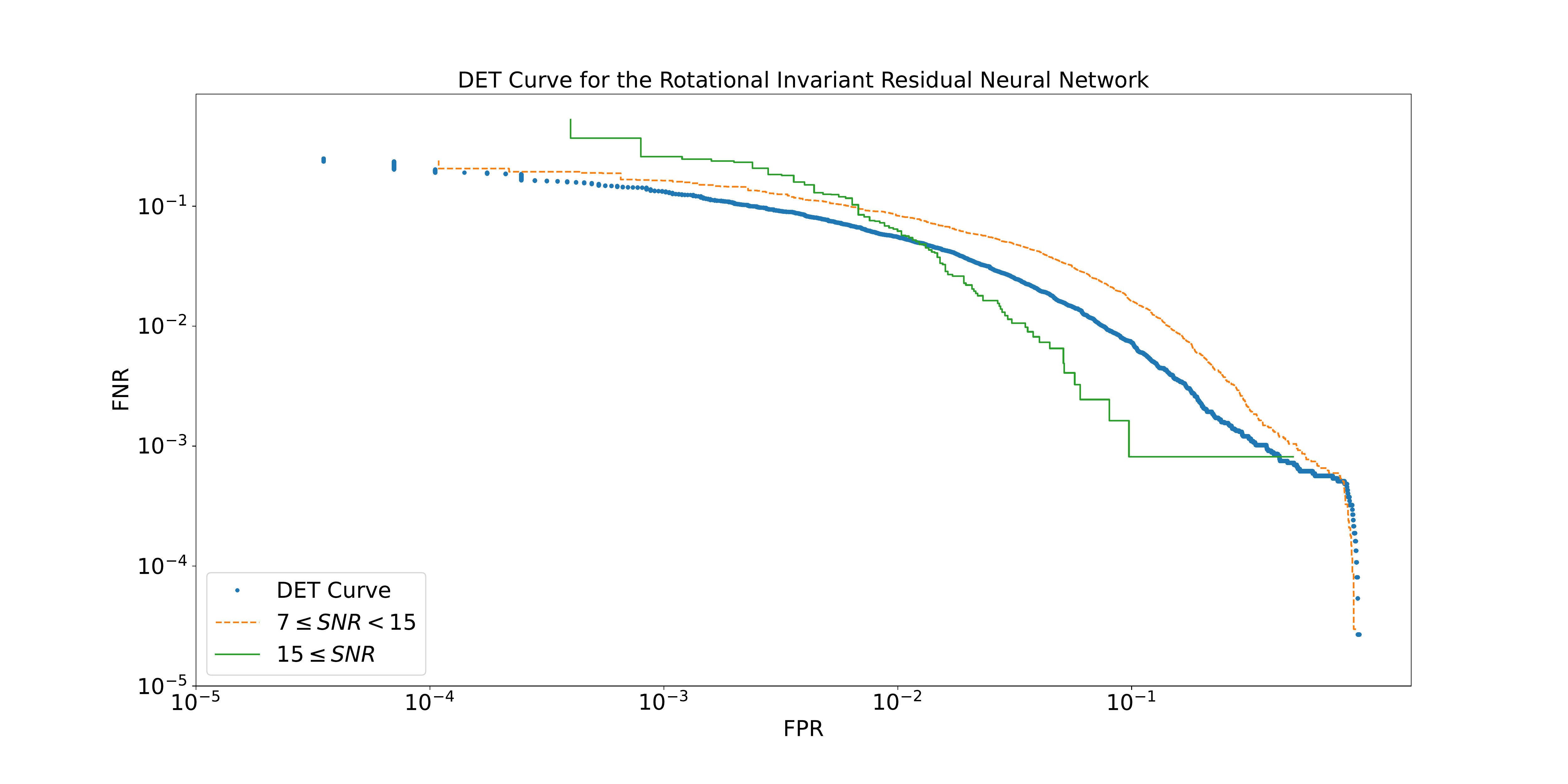}
\end{center}

\caption{The blue dots show the Detection Error Trade-off curve for the stamps from 3$\times$3 binned images. The green and orange curves show the DET curve for different signal-to-noise ratio groups. The x-axis is the false-negative rate, and the y-axis is the false positive rate.}\label{fig:det}
\end{figure}

We calculated the false-negative rate (FNR) and the false positive rate (FPR) with scikit-learn to analyse the neural network's performance. 
The detection error trade-off (DET) curve demonstrates how FNR is correlated and FPR. The DET curves could show the performance of the CNN models used for candidate classification and provide direct feedback on the detection error trade-offs to help analyse the neural network. In \autoref{fig:det}, we present the DET curve for candidates in different signal-to-noise (SNR) groups. The higher SNR curve shows a quicker move to the bottom left, better fitting the plot.

\autoref{fig:det} shows that the rotational invariant model is well-operated for the higher signal-to-noise ratio sources, which is in line with our expectations. The curve for higher SNR shows a vertical line, which may be caused by having only $10^4$ sources as positive stamps. 

\section{Conclusion}
\label{sec:conclusion}
This paper describes the science data reduction pipeline and transient detection pipeline for the AST3-3 telescope at the Yaoan Observation Station. The science image pipeline uses the statistical method to estimate the quality of the observed image, taking into account the effects of poor weather, such as clouds passing through in the FoV.

The transient detection pipeline uses multiple binned and unbinned science images to extract the candidates faster and deeper. In terms of transient source detection, the robustness and flexibility of the program are improved through a combination of multiple detection methods and the CNN method.  
We introduce a rotation-invariant residual neural network to classify the candidate stamps from the transient detection pipeline. The CNN trained on the negative and simulated positive stamps cut. 
The CNN accuracy achieved 99.87\% for the QIP and 99.20\% for the DIP. AST3-3 has been designed for robotic observation and has a complete pipeline system with specific software, a well-trained CNN model, and management for the observation at Yaoan Observation Station. This work allows us to effectively participate in the LIGO-Virgo O4 ground optical follow-up observing campaign.

\section*{Funding}
This work is partially supported by the National Natural Science Foundation of China (grant Nos. 11725314, 12041306), the Major Science and Technology Project of Qinghai Province (2019-ZJ-A10), ACAMAR Postdoctoral Fellow, China Postdoctoral Science Foundation (Grant No. 2020M681758) and Natural Science Foundation of Jiangsu Province (grant No. BK20210998). Tianrui Sun and Alberto Castro-Tirado also acknowledge financial support from the State Agency for Research of the Spanish MCIU through the "Center of Excellence Severo Ochoa" award to the Instituto de Astrofísica de Andalucía (SEV-2017-0709). Tianrui Sun thanks the China Scholarship Council (CSC) for funding his PhD scholarship (202006340174).  

\section*{Acknowledgments}
The AST3-3 team would like to express their sincere thanks to the staff of the Yaoan Observation Station. The authors are also grateful to anonymous referees whose opinion has signiﬁcantly improved this manuscript.
This research has made use of data and services provided by the International Astronomical Union's Minor Planet Center.
This work has made use of data from the European Space Agency (ESA) mission
{\it Gaia} (\url{https://www.cosmos.esa.int/gaia}), processed by the {\it Gaia}
Data Processing and Analysis Consortium (DPAC,
\url{https://www.cosmos.esa.int/web/gaia/dpac/consortium}). Funding for the DPAC has been provided by national institutions, in particular the institutions participating in the {\it Gaia} Multilateral Agreement.

Software packages: This research made use of Astropy,\footnote{http://www.astropy.org} a community-developed core Python package for Astronomy \citep{astropy:2013, astropy:2018}. The python packages: L.A. Cosmic \citep{2012ascl.soft07005V}, pyephem \citep{2011ascl.soft12014R}, skyfield \citep{2019ascl.soft07024R}, Pytorch \citep{NEURIPS2019_9015}, scikit-Image \citep{van2014scikit}, scikit-learn \citep{developers2018scikitlearn}, matplotlib \citep{matplotlib}, scipy\citep{jones2001scipy}, statsmodels \citep{seabold2010statsmodels}, sep \citep{2016JOSS....1...58B}, SFFT \citep{sfft_zenodo}. Software programs: SExtractor \citep{2010ascl.soft10064B}, SWarp \citep{2010ascl.soft10068B}, Astrometry.net \citep{2012ascl.soft08001L}, FITSH \citep{2011ascl.soft11014P}, fpack \citep{2010ascl.soft10002S}. 

\bibliographystyle{frontiersinSCNS_ENG_HUMS} 
\bibliography{test}

\end{document}